\documentclass[preprint2]{emulateapj}
\usepackage{natbib}
\bibpunct[; ]{(}{)}{,}{a}{}{;}

\shorttitle{Spatially-Resolved Nonthermal Line Broadening During the Impulsive Phase of a Solar Flare}

\shortauthors{R. O. Milligan}

\begin{document}

\title{Spatially-Resolved Nonthermal Line Broadening During the Impulsive Phase of a Solar Flare}

\notetoeditor{the contact email is r.milligan@qub.ac.uk and is the only one which should
appear on the journal version}

\author{Ryan O. Milligan\altaffilmark{1,*}}

\altaffiltext{1}{Solar Physics Laboratory (Code 671), Heliophysics Science Division, NASA Goddard Space Flight Center, Greenbelt, MD 20771, U.S.A.}
\altaffiltext{*}{Current address: Astrophysics Research Centre, School of Mathematics \& Physics, Queen's University Belfast, University Road, Belfast, Northern Ireland, BT7 1NN}

\begin{abstract}
This paper presents a detailed study of excess line broadening in EUV emission lines during the impulsive phase of a C-class solar flare. In this work, which utilizes data from the EUV Imaging Spectrometer (EIS) onboard Hinode, the broadened line profiles were observed to be co-spatial with the two HXR footpoints as observed by RHESSI. By plotting the derived nonthermal velocity for each pixel within the \ion{Fe}{15} and \ion{Fe}{16} rasters against its corresponding Doppler velocity a strong correlation ($\vert r \vert > 0.59$) was found between the two parameters for one of the footpoints. This suggested that the excess broadening at these temperatures is due to a superposition of flows (turbulence), presumably as a result of chromospheric evaporation due to nonthermal electrons. Also presented are diagnostics of electron densities using five pairs of density-sensitive line ratios. Density maps derived using the \ion{Mg}{7} and \ion{Si}{10} line pairs showed no appreciable increase in electron density at the footpoints, while the \ion{Fe}{12}, \ion{Fe}{13}, and \ion{Fe}{14} line pairs revealed densities approaching 10$^{11.5}$~cm$^{-3}$. Using this information, the nonthermal velocities derived from the widths of the two \ion{Fe}{14} lines were plotted against their corresponding density values derived from their ratio. This showed that pixels with large nonthermal velocities were associated with pixels of moderately higher densities. This suggests that nonthermal broadening at these temperatures may have been due to enhanced densities at the footpoints, although estimates of the amount of opacity broadening and pressure broadening appeared to be negligible.
\end{abstract}


\section{INTRODUCTION}
\label{intro}

The spectroscopy of extreme ultra-violet (EUV) emission lines is a crucial diagnostic tool for determining the composition and dynamics of the flaring solar atmosphere. While imaging instruments provide important context information of the morphology and structure of coronal features, the images themselves are usually broadband, comprising several different ion species which can bias the interpretation of the observations. Spectroscopy offers the advantage of providing quantifiable measurements of parameters such as temperature, density, and velocity, which can then be compared with predictions from theoretical models. 

In the context of solar flares, EUV and soft X-ray (SXR) spectroscopy has led to important measurements of chromospheric evaporation through Doppler shifts of high-temperature line profiles. \cite{acto82}, \cite{anto83}, \cite{canf87}, \cite{zarr88}, and \cite{dosc05} each measured blueshifts of 300--400~km~s$^{-1}$ in the \ion{Ca}{19} line (3.1--3.2~\AA, 25 MK) using the Bent and Bragg Crystal Spectrometers (BCS) onboard SMM \citep{acto81} and Yohkoh \citep{culh91}, respectively. Similar studies using data from the Coronal Diagnostic Spectrometer (CDS; \citealt{harr95}) on SOHO revealed upflow velocities of 150--300~km~s$^{-1}$ in the \ion{Fe}{19} line (592.23~\AA, 8 MK; \citealt{czay99,czay01,teri03,bros04,mill06a,mill06b,bros07,bros09a,bros09b}). The EUV Imaging Spectrometer (EIS) onboard Hinode now allows these measurements to be made over many high temperature lines simultaneously \citep{mill09,delz11,grah11}, and its superior spectral resolution, coupled with its imaging capability now means that spatial information regarding line widths can be obtained; something not previously possible with other instruments.

The width of spectral lines reveals important information on the temperature and turbulence of the emitting plasma. Line width is generally made up of at least three components: the intrinsic instrumental resolution, the thermal Doppler width, and any excess (nonthermal) broadening which can be an indicator of possible turbulence, pressure or opacity broadening, or the Stark Effect. Many studies have reported excess EUV and SXR line broadening, over and above that expected from thermal emission, during a flare's impulsive phase indicating possible turbulent motion. This was typically observed in the \ion{Ca}{19} resonance line (100--130~km~s$^{-1}$; \citealt{dosc80,feld80,gabr81,anto82}) and the \ion{Fe}{25} line (1.85~\AA, 90~km~s$^{-1}$; \citealt{grin73}), although this emission was integrated over the entire disk. Opacity effects have been observed in stellar flare spectra, in particular in \ion{C}{3} lines, although no actual opacity broadening was conclusively measured \citep{chri04,chri06}. The effect of Stark broadening due to the electrostatic field of the charged particles in the plasma has been studied extensively in the Balmer series of hydrogen (e.g. \citealt{lee96}) and in stellar flare spectra \citep{john97}. \cite{canf84} also noted that the excess emission in the wings of the H$\alpha$ line was critically dependent on the flux of the incident electrons during solar flares. 

The origin of excess broadening of optically thin emission lines beyond their thermal Doppler widths, even in quiescent active region spectra, is still not fully understood \citep{dosc08,imad08}. The general consensus is that the broadening is due to a continuous distribution of different plasma flow speeds in structures smaller than the spatial resolution of the spectrometer \citep{dosc08}. Several studies have been carried out which correlate Doppler velocity with nonthermal velocity for entire active regions using raster data from EIS \citep{hara08,dosc08,brya10,pete10}. Each of these studies showed that Doppler speed and nonthermal velocities were well correlated over a given quiescent active region indicating that the broadening is likely due to a distribution of flow speeds. However excess line broadening could also be due pressure broadening resulting from increased electron densities. In these cases, collisions with electrons occur on time scales shorter than the emission time scale of the ion, resulting in a change in frequency of the emitted photon. However, \cite{dosc07} found that regions of high temperature in an active region corresponded to regions of high densities, but the locations of increased line width did not, suggesting that pressure broadening was not the correct explanation in this instance. Also using EIS, \cite{hara09} suggested that turbulence in the corona could be induced by shocks emanating from the reconnection site.

EIS also offers the ability to obtain values of the coronal electron density by taking the ratio of the flux of two emission lines from the same ionization stage when one of the lines is derived from a metastable transition. \cite{gall01} and \cite{mill05} used various coronal line ratios from SOHO/CDS data to determine the density structure of active regions. \cite{warr03} used the Solar Ultraviolet Measurements of Emitted Radiation (SUMER) spectrometer, also on SOHO, to determine the density structure of an active region above the limb. More recently, several similar studies have been made using the density diagnostic capabilities of EIS. As mentioned above, \cite{dosc07} found that regions of high temperature in an active region corresponded to regions of high densities, but the locations of increased line width did not. \cite{chif08} determined the density in upflowing \ion{Fe}{12} material in a jet and found that the faster moving plasma was more dense. More recently \cite{grah11} found enhanced electron densities from \ion{Fe}{12}, \ion{Fe}{13}, and \ion{Fe}{14} ratios at a flare footpoint.

\begin{figure}[!t]
\begin{center}
\includegraphics[width=8.5cm]{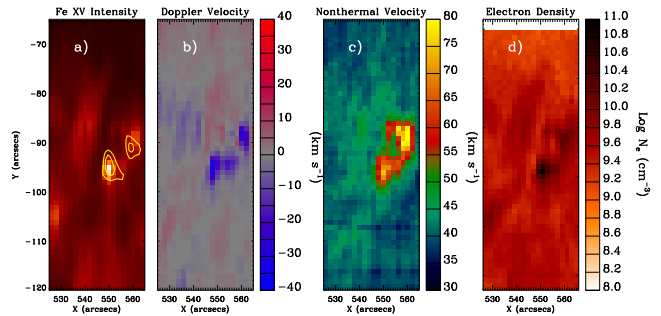}
\caption{Derived plasma parameters from a single EIS raster taken during the impulsive phase of a C1.1 flare that occurred on 2007 December 14. a) A image showing the spatial distribution of the \ion{Fe}{15} 284.16\AA~line intensity. Overlaid are the contours of the 20-25~keV X-ray sources as observed by RHESSI. b) The corresponding Doppler velocity map derived from shifts in the line centroid relative to a quiet-Sun value. Positive velocities (redshifts) indicate downflows, while negative velocities (blueshifts) indicate upflows. c) Map of the nonthermal velocity from the line widths over and above the thermal plus instrumental widths. d) Spatial distribution of electron density from the ratio of two \ion{Fe}{14} lines (264.79\AA/274.20\AA) which are formed at a similar temperature to that of \ion{Fe}{15}.}
\label{fe15_int_vel_den}
\end{center}
\end{figure}

This paper continues the work of \cite{mill09}, which focused primarily on measuring the Doppler shifts of 15 EUV emission lines covering the temperature range 0.05--16 MK during the impulsive phase of a C-class flare that occurred on 2007 December 14. In doing so, a linear relationship was found between the blueshift of a given line and the temperature at which it was formed. The work also revealed the presence of redshifted footpoint emission (interpreted as chromospheric condensation due to the overpressure of the evaporating material), at temperatures approaching 1.5~MK; much higher than predicted by current solar flare models (see also \citealt{mill08}). During the initial analysis of the EIS data from this event, it was noticed that the EUV line profiles at the location of the hard X-ray (HXR) emission were broadened beyond their thermal width in addition to being shifted from their `rest' wavelengths. Furthermore, the corresponding electron density maps yielded substantially high density values ($\ge$10$^{10}$~cm$^{-3}$) at the same location. Figure~\ref{fe15_int_vel_den} shows a sample of data products derived from the \ion{Fe}{15} 284.16\AA~raster taken during the impulsive phase: an intensity map ($a$; with contours of the 20--25~keV emission observed by RHESSI overlaid), a Doppler map ($b$), a nonthermal velocity map ($c$), and a density map ($d$; derived from the \ion{Fe}{14} line ratio (264.79\AA/274.20\AA) which is formed at a similar temperature). At the location of the HXR emission, the plasma appeared to be blueshifted, turbulent, and dense. This then raised the question: `what was the nature of the nonthermal line broadening at the site of the HXR emission during the impulsive phase of this solar flare?' Was it due to unresolved plasma flows similar to that found in active region studies \citep{hara08,dosc08,brya10,pete10} or was it from pressure or opacity broadening due to high electron densities similar to that found in optically thick H lines \citep{canf84,lee96,chri04,chri06}? 

Thanks to the rich datasets provided by EIS during this event, a much more comprehensive analysis of the flaring chromosphere can be carried out. The observing sequence that was running during this event contained over 40 emission lines (including 5 density sensitive pairs) and rastered over the flaring region with a cadence of 3.5 minutes. This allowed measurements of differential emission measure (from line intensities), Doppler velocity (from line shifts), thermal and nonthermal broadening (from line widths), and electron densities (from line ratios) over the same broad temperature range covered by \cite{mill09} to be made. Section~\ref{eis_obs} presents a brief overview of the event. Section~\ref{line_fit_vel_anal} describes the derivation of the various plasma parameters. Section~\ref{results} discusses the findings from correlative studies between parameters while the conclusions are presented in Section~\ref{conc}.

\begin{figure}[!t]
\begin{center}
\includegraphics[width=8.5cm]{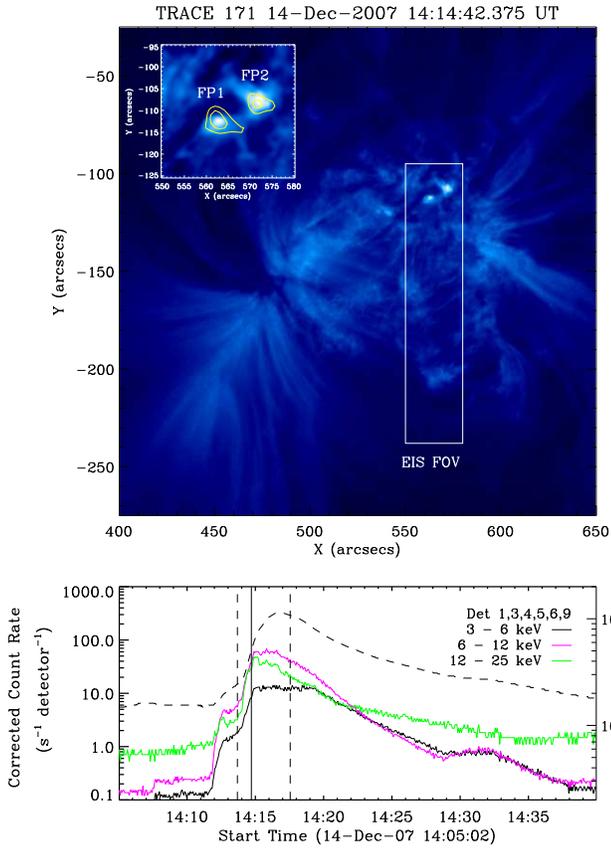}
\caption{Top: An image of NOAA AR 10978 taken in the TRACE 171~\AA~passband on 2007 December 14 at 14:14:42~UT. Overlaid is the rectangular field of view of the EIS raster. The inset in the top left corner shows a zoomed-in portion of the image containing the two HXR footpoints (FP1 and FP2) under investigation. The contours overlaid in yellow are the 60\% and 80\% levels of the 20--25~keV emission as observed by RHESSI from 14:14:28--14:15:00~UT. Bottom: Lightcurves in the 3--6 (black), 6--12 (magenta), and 12--15~keV (green) energy bands from RHESSI. The dashed lightcurve indicates the corresponding 1--8~\AA~ emission from GOES. The vertical dashed lines denote the start and end times of the EIS raster taken during the impulsive phase, while the vertical solid line marks the time of the TRACE and RHESSI images in the top panel.}
\label{trace_hsi_eis_fov}
\end{center}
\end{figure}

\section{The 2007 December 14 Flare}
\label{eis_obs}

The GOES C1.1 class flare under study occurred in NOAA AR 10978 on 2007 December 14 at 14:12 UT. The top panel of Figure~\ref{trace_hsi_eis_fov} shows an image of the active region taken by the Transition Region and Coronal Explorer (TRACE; \citealt{hand99}) in the 171~\AA~passband during the impulsive phase of the flare. Two bright EUV footpoints are visible in the northern end of the box which denotes the EIS field of view (FOV). The inset in the top left corner of the panel shows a close-up of the footpoints with contours of the 20--25~keV emission observed by the Ramaty High-Energy Solar Spectroscopic Imager (RHESSI; \citealt{lin02}) overlaid. After manually correcting for the 5$\arcsec$ pointing offset in both the solar X and solar Y directions, the two EUV footpoints align well with the HXR sources as seen by RHESSI, here labelled as FP1 and FP2. The bottom panel of the figure shows the X-ray lightcurves from RHESSI in the 3--6, 6--12, and 12--25~keV energy bands, along with the 1--8~\AA~lightcurve from GOES. The vertical solid line denotes the time of the TRACE and RHESSI images in the top panel, while the vertical dashed lines mark the start and end times of the EIS raster under investigation.

The observing study that EIS was running when the flare occurred (CAM\_ARTB\_RHESSI\_b\_2) was originally designed to search for active region and transition region brightenings in conjunction with RHESSI. Using the 2$\arcsec$ slit, EIS rastered across a region of the Sun, from west to east, covering an area of 40$\arcsec \times$143$\arcsec$, denoted by the rectangular box in Figure~\ref{trace_hsi_eis_fov}. Each slit position had an exposure time of 10~s resulting in an effective raster cadence of $\sim$3.5~minutes. These fast-raster studies are preferred for studying temporal variations of flare parameters while preserving the spatial information. Equally important though, is the large number of emission lines which covered a broad range or temperatures. This observing study used 21 spectral windows, some of which contain several individual lines. The work presented here focuses on 15 lines spanning the temperature range 0.05--16~MK. Details of the lines, their rest wavelengths and peak formation temperatures are given in Table~\ref{line_data}, along with their Doppler velocities derived by \cite{mill09}\footnote[1]{Note that \cite{mill09} originally used formation temperatures quoted in \cite{youn07} whereas this work used revised values from the latest version of CHIANTI (v6.0.1; \citealt{dere09}). Also, the \ion{Fe}{13} line was incorrectly identified as being redshifted in the original analysis. The followup analysis presented here revealed it to be blueshift and the revised velocity is quoted in Table~\ref{line_data}.} and the nonthermal velocities as measured in this work. The majority of these lines are well resolved and do not contain blends, thereby reducing ambiguities in their interpretation. 

\begin{figure*}
\begin{center}
\includegraphics[height=24cm,angle=180]{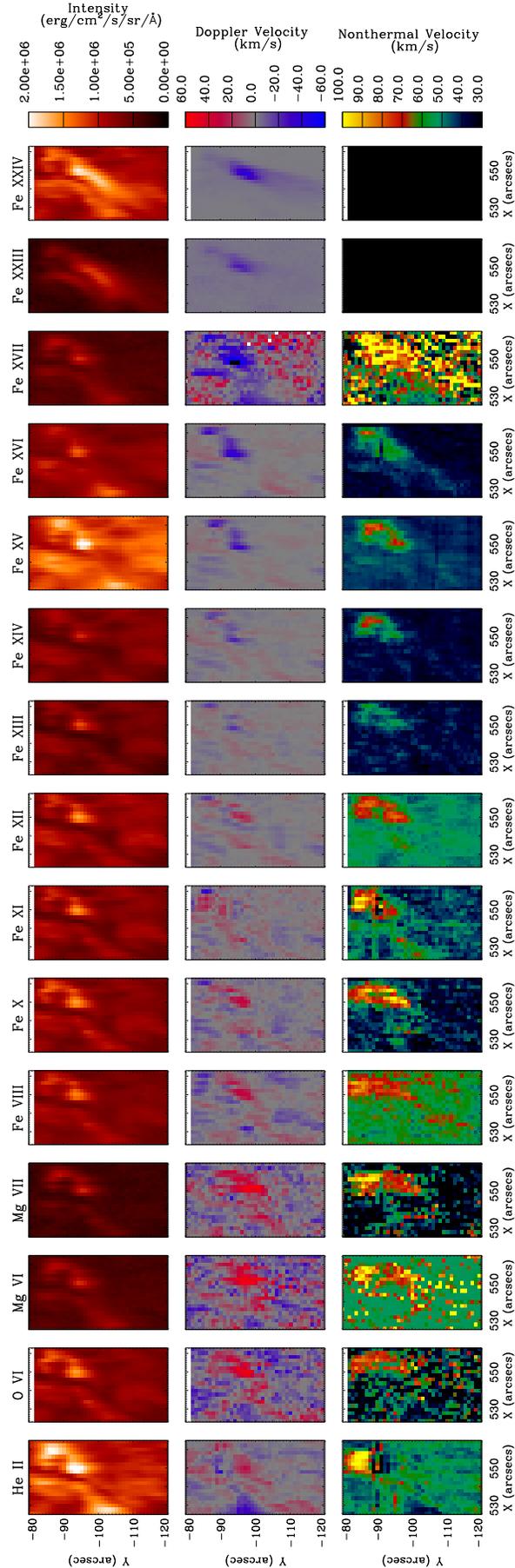}
\caption{Partial field-of-view of the EIS raster taken during the impulsive phase of the flare in each of the 15 emission lines used in this study. Top row shows the normalized intensity maps. The middle and bottom rows show the corresponding Doppler velocity and nonthermal velocity maps, respectively. In the Doppler maps, positive velocities (redshifts) indicate downflows, while negative velocities (blueshifts) indicate upflows. }
\label{eis_int_vel_wid_maps}
\end{center}
\end{figure*}

\begin{table}
\begin{center}
\small
\caption{\textsc{\small{Ions, Wavelengths, And Peak Formation Temperatures Of Emission Lines Used In This Work Along With Measured Doppler And Nonthermal Velocities.}}}
\label{line_data}
\begin{tabular}{lcccc} 
\tableline
\tableline
\multicolumn{1}{c}{Ion}	&$\lambda$(\AA) &$T$ (MK)\footnote[1]{Line formation temperatures from CHIANTI v6.0.1. \cite{dere09}. Note that these differ slightly from those quoted in \cite{mill09} which were from \cite{youn07}.} 	&$v$ (km~s$^{-1}$)\footnote[2]{From \cite{mill09}.} &$v_{nth}$ (km~s$^{-1}$)\\ 
\tableline
\ion{He}{2}		&256.32	&0.05	&21$\pm$12		&57\\
\ion{O}{6}			&184.12	&0.3		&60$\pm$14		&68\\
\ion{Mg}{6}		&268.99	&0.5		&51$\pm$15		&71\\
\ion{Mg}{7}		&280.75	&0.6		&53$\pm$13		&64\\
\ion{Fe}{8}		&185.21	&0.8		&33$\pm$17		&74\\
\ion{Fe}{10}		&184.54	&1.0		&35$\pm$16		&97\\
\ion{Fe}{11}		&188.23	&1.2		&43$\pm$15		&60\\
\ion{Fe}{12}		&195.12	&1.35	&28$\pm$17		&81\\
\ion{Fe}{13}\footnote[3]{Note that in \cite{mill09} this line was incorrectly identified as being redshifted.}
				&202.04 	&1.6		&-18$\pm$14		&54\\
\ion{Fe}{14}		&274.20	&1.8		&-22$\pm$12		&58\\
\ion{Fe}{15}		&284.16	&2.0		&-32$\pm$8		&73\\
\ion{Fe}{16}		&262.98	&2.5		&-39$\pm$20		&48\\
\ion{Fe}{17}		&269.17	&4.0		&-69$\pm$18		&78\\
\ion{Fe}{23}		&263.76	&14.0	&$<$-230$\pm$32	&122\\
\ion{Fe}{24}		&192.03	&18.0	&$<$-257$\pm$28	&105\\
\tableline
\normalsize
\end{tabular}
\end{center}
\end{table}

Intensity, Doppler, and nonthermal velocity maps in each of the 15 emission lines are shown in Figure~\ref{eis_int_vel_wid_maps} for the portion of the EIS raster containing the two footpoints during the impulsive phase of the flare. Looking at the brighter southeastern footpoint in the top  row of Figure~\ref{eis_int_vel_wid_maps}, there are no discernible differences between images formed at temperatures lower than $\sim$4~MK. Images in the two hottest lines (\ion{Fe}{23} and \ion{Fe}{24}) however, show an overlying loop structure which had begun to fill with hot plasma. For a more detailed description of this event, see \cite{mill09}.

\section{Data Analysis}
\label{line_fit_vel_anal}

\subsection{Doppler and Nonthermal Velocities}
\label{velocities}

Each line profile in each pixel within a raster was fitted with a single Gaussian profile. The Doppler and nonthermal velocities were calculated from the line centroids and line widths, respectively. The line of sight component to the Doppler velocity, $v$, is given by:

\begin{equation}
\frac{v}{c}= \frac{\lambda - \lambda_0}{\lambda_0}
\end{equation}

\noindent
where $\lambda$ is the measured line centroid, $\lambda_0$ is the reference (rest) wavelength obtained from quiet-Sun values (except for the \ion{Fe}{23} and \ion{Fe}{24} lines which were measured relative to centroid positions taken during the flare's decay phase), and $c$ is the speed of light. The resulting Doppler velocity maps for each of the 15 lines are shown in the middle row of Figure~\ref{eis_int_vel_wid_maps}. This shows that emission from lines formed below $\sim$1.35~MK was redshifted at the loop footpoints while plasma at higher temperatures (2--16~MK) was blueshifted (from \citealt{mill09}). 

The nonthermal velocity, $v_{nth}$, can be calculated using:

\begin{equation}
W^2 = 4 ln2 \left(\frac{\lambda}{c}\right)(v_{th}^{2} + v_{nth}^{2}) + W_{inst}^{2}
\end{equation}

\noindent
where $W$ is the measured width of the line profile, and $W_{inst}$ is the instrumental width (taken here to be 0.056 m\AA~from \citealt{dosc07} and \citealt{harr09}). The thermal velocity, $v_{th}$, is given by:

\begin{equation}
\sqrt\frac{2k_{B}T}{M}
\label{eqn:therm_vel}
\end{equation}

\noindent
where $k_B$ is the Boltzmann constant, $T$ is the formation temperature of the line, and $M$ is the mass of the ion. The resulting nonthermal velocity maps are shown in the bottom row of Figure~\ref{eis_int_vel_wid_maps}. From this it can be seen that nearly all lines exhibit some degree of broadening at the loop footpoints, although some maps appear `noisier' than others. This was particularly true for the \ion{Fe}{23} and \ion{Fe}{24} lines (not shown) which have no quiet-Sun emission. Furthermore, as noticed in \cite{mill09}, the line profiles at the flare footpoints for these ions also required a two-component fit (one stationery, one blueshifted) with the blueshifted component extending beyond the edge of the spectral window in many cases, further complicating the construction of a nonthermal velocity map.

\subsection{Density Diagnostics and Column Depths}
\label{density}

\begin{table}
\begin{center}
\small
\caption{\textsc{\small{Ions, Wavelengths, And Peak Formation Temperatures Of Density Sensitive Line Ratios Used In This Work Along With The Range Of Densities Over Which They Are Sensitive.}}}
\label{density_lines}
\begin{tabular}{lccc} 
\tableline
\tableline
\multicolumn{1}{c}{Ion}	&$\lambda$(\AA) &$T$ (MK)		&$n_e$ (cm$^{-3}$)\\ 
\tableline
\ion{Mg}{7}	&278.40	&0.6		&10$^{8}$--10$^{10}$	\\
\ion{Mg}{7}	&280.75	&0.6		&10$^{8}$--10$^{10}$	\\
\ion{Fe}{12}	&195.12	&1.35	&10$^{7}$--10$^{11}$	\\
\ion{Fe}{12}	&196.64	&1.35	&10$^{7}$--10$^{11}$	\\
\ion{Si}{10}	&258.37	&1.4		&10$^{8}$--10$^{9}$	\\
\ion{Si}{10}	&261.04	&1.4		&10$^{8}$--10$^{9}$	\\
\ion{Fe}{13}	&202.04 	&1.6		&10$^{7}$--10$^{10}$	\\
\ion{Fe}{13}	&203.83 	&1.6		&10$^{7}$--10$^{10}$	\\
\ion{Fe}{14}	&264.79	&1.8		&10$^{9}$--10$^{11}$	\\
\ion{Fe}{14}	&274.20	&1.8		&10$^{9}$--10$^{11}$	\\
\tableline
\normalsize
\end{tabular}
\end{center}
\end{table}

\begin{figure}
\begin{center}
\includegraphics[height=8.5cm,angle=90]{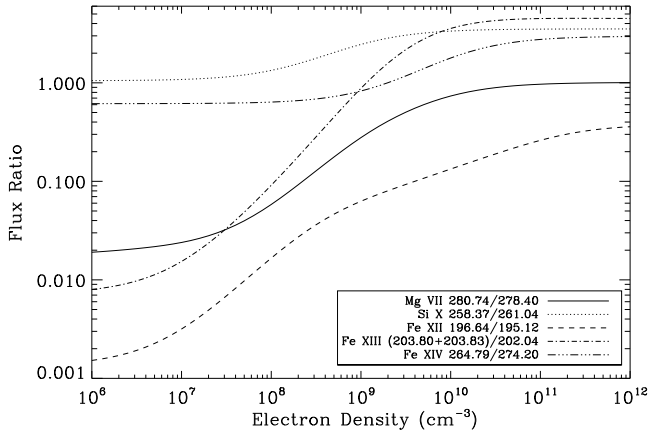}
\caption{The theoretical relationships between line flux and derived electron density from CHIANTI v6.0.1 for each of the 5 line pairs used in this study.}
\label{plot_eis_chianti_den_ratios}
\end{center}
\end{figure}

\begin{figure}[!t]
\begin{center}
\includegraphics[width=8.5cm,angle=90]{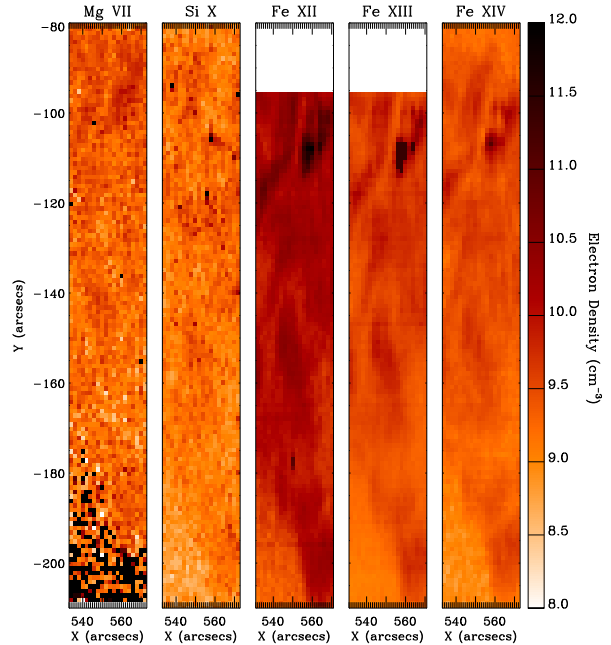}
\caption{Electron density maps in each of the 5 line pairs available in this study. The ``missing data'' at the top of the \ion{Fe}{12} and \ion{Fe}{13} rasters are due to the 17$\arcsec$ offset (in the $y$-direction) between the two EIS detectors.}
\label{eis_density_plot_5_lines}
\end{center}
\end{figure}

The EIS dataset used in this work contained five pairs of density sensitive line ratios: \ion{Mg}{7}, \ion{Si}{10}, \ion{Fe}{12}, \ion{Fe}{13}, and \ion{Fe}{14} (see Table~\ref{density_lines} for details). The theoretical relationship between the flux ratios and the corresponding electron densities as derived from CHIANTI v6.0.1 are shown in Figure~\ref{plot_eis_chianti_den_ratios}. Each of these line pairs are mostly sensitive to densities in the range $\sim$10$^{8}$--10$^{10}$~cm$^{-3}$. Using the {\sc eis\_density.pro} routine in SSWIDL, electron density maps were compiled for the raster taken during the impulsive phase at each of these five temperatures. These maps are shown in Figure~\ref{eis_density_plot_5_lines}. Both the maps formed from \ion{Mg}{7} and \ion{Si}{10} line pairs show no discernible evidence for enhanced densities at the location of the HXR emission. As the \ion{Mg}{7} lines are formed at temperatures corresponding to the lower transition region, where densities are already on the order of 10$^{10}$~cm$^{-3}$, any appreciable increase would be difficult to detect. Similarly, the \ion{Si}{10} lines are only sensitive to densities below 10$^{9}$~cm$^{-3}$ (from Table~\ref{density_lines} and Figure~\ref{plot_eis_chianti_den_ratios}) and may therefore not be suitable for measuring density enhancements during flares. The \ion{Fe}{12} map, while showing enhanced densities at the loop footpoints relative to the quiet Sun, exhibits a systematically higher density value (by approximately a factor of 2) than either the \ion{Fe}{13} and \ion{Fe}{14} maps, which are formed at comparable temperatures. This discrepancy is likely due to inaccuracies in the atomic data for \ion{Fe}{12} rather than a real, physical difference in the densities sampled by the different ions (P. Young; priv. comm. See also \citealt{youn09} and \citealt{grah11}). The \ion{Fe}{13} and \ion{Fe}{14} maps themselves show a distinct increase in electron densities at the loop footpoints with the values from the \ion{Fe}{13} pair reaching their high density limits.

Using the values derived for the electron densities it is possible to compute the column depth of the emitting material. Given that the intensity of a given emission line, $I$, can be expressed as:

\begin{equation}
4 \pi I = 0.83 \int G(T, N_{e}) N_{e}^{2}dh
\label{col_depth_one}
\end{equation}

\noindent
where $G(T,N_{e})$ is the contribution function for a given line, $N_{e}$ is the electron number density and $h$ is the column depth. By approximating the contribution function as a step function around $T_{max}$ and assuming that the density is constant across each pixel, Equation~\ref{col_depth_one} can be written as:

\begin{equation}
4 \pi I = 0.83 G_{0} N_{e}^{2} h
\end{equation}

\begin{figure}[]
\begin{center}
\includegraphics[width=8.5cm,angle=90]{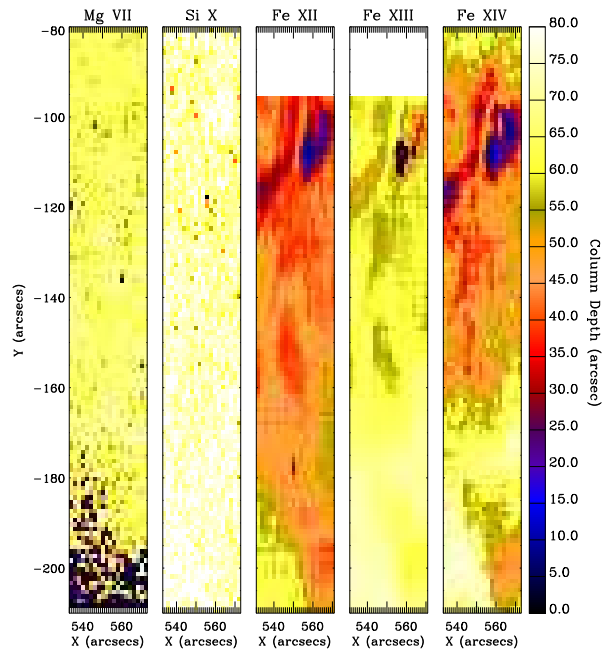}
\caption{Column depth maps (in arcseconds) in each of the 5 density sensitive line pairs available in this study.}
\label{eis_col_depth_plot_5_lines}
\end{center}
\end{figure}

\noindent
The  {\sc eis\_density.pro} routines calculates $G_{0}$ for a given electron density which allows the value of $h$ to be derived for each pixel within a raster for which the density is known (see \citealt{youn11} for more details). Figure~\ref{eis_col_depth_plot_5_lines} shows the maps of column depth for the five density maps displayed in Figure~\ref{eis_density_plot_5_lines}. Unsurprisingly, the spatial distribution of column depth closely resembles that of the density distributions, with footpoint emission exhibiting smaller column depths than the surrounding active region; less than 15$\arcsec$ in most cases, and as little as 0.01$\arcsec$ in some places. These values agree well with those found by \cite{delz11}, who used the same technique and line ratio but assumed photospheric abundances rather than coronal, and with \cite{sain10} who derived column depth estimates from RHESSI HXR observations. Information on the column depths can be used to determine the opacity at the footpoints during this event. This will be discussed further in Section~\ref{den_nth_vel}. 

\section{Results}
\label{results}

\begin{figure}
\begin{center}
\includegraphics[width=8.5cm]{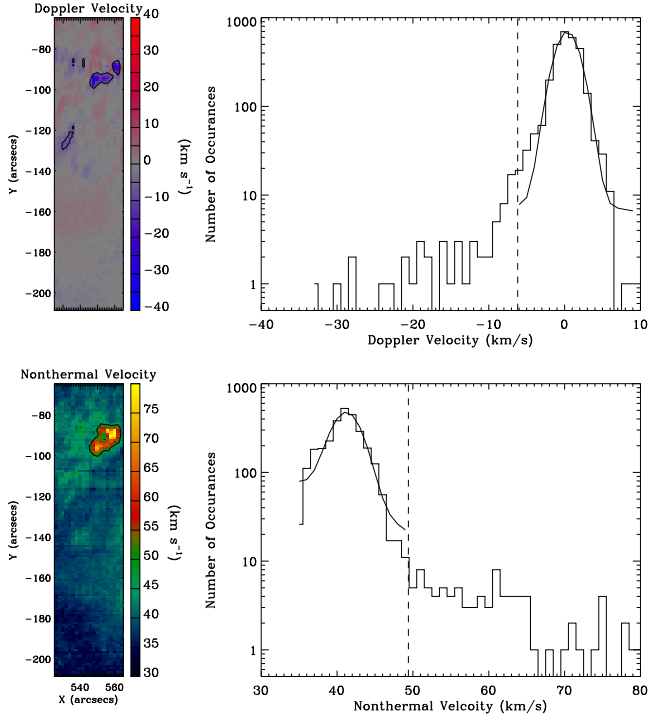}
\caption{Top row: A velocity map of the entire EIS raster in the \ion{Fe}{15} line taken during the impulsive phase, and the corresponding histogram of Doppler velocity values. Bottom row: The nonthermal velocity map for the same raster and the corresponding histogram of nonthermal velocity values. The solid curves on each of the histogram plots are Gaussian fits to the distributions. The vertical dashed lines mark the 3$\sigma$ width of the Gaussians, which are then overlaid as contours on the maps. This 3$\sigma$ level adequately differentiates the flaring footpoint emission from the rest of the active region.}
\label{plot_fe_xv_vel_vnth_hist}
\end{center}
\end{figure}

Previous studies of active region heating using EIS data have attempted to establish the cause of line broadening by correlating the Doppler velocity at each pixel in a raster with its corresponding nonthermal velocity as determined from the line width. The same method was applied to the data in this work to explore the possible mechanisms for line broadening at the footpoints of a flaring loop. In order to distinguish flaring emission from that of the surrounding active region and quiet-Sun plasma, histograms of all data values were plotted. Figure~\ref{plot_fe_xv_vel_vnth_hist} shows the Doppler and nonthermal velocity maps and corresponding histograms for the Fe XV line during the impulsive phase. In both cases, the distribution of values is close to Gaussian (centered on zero km~s$^{-1}$ in the Doppler velocity case and on $\sim$41~km~s$^{-1}$ in the nonthermal velocity case). Data values that lay outside the 3$\sigma$ level of the Gaussian fit to the histograms were found to correspond to emission coming solely from the footpoints as illustrated by the contours overplotted on the maps (i.e. the contours drawn correspond to the 3$\sigma$ level of the Gaussian fit in each case). This was repeated for the \ion{Fe}{14} and \ion{Fe}{16} lines which had the strongest signal-to-noise ratios as well as appreciable Doppler velocities.

\begin{figure*}[!t]
\begin{center}
\includegraphics[height=16cm,angle=90]{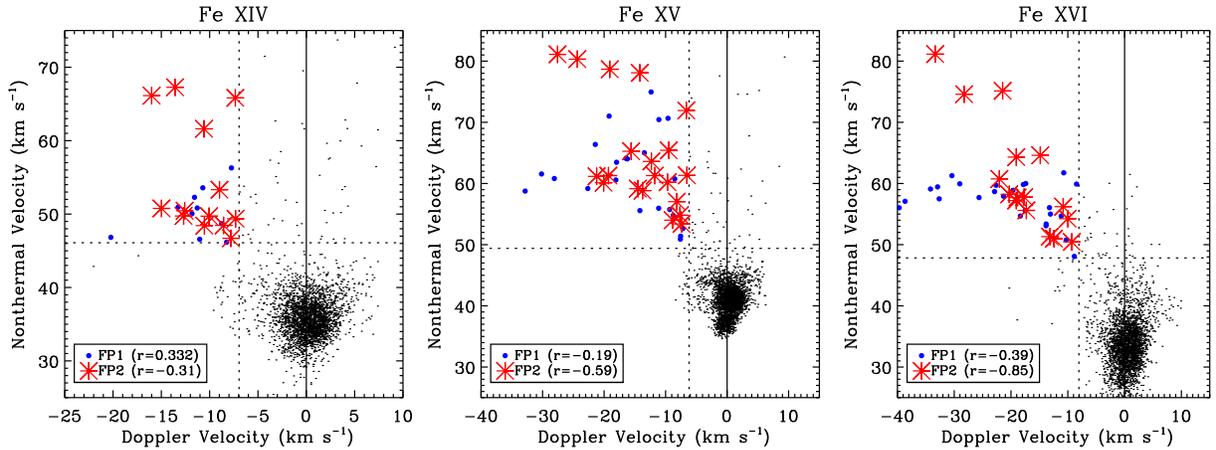}
\caption{Scatter plots of Doppler velocity against nonthermal velocity for \ion{Fe}{14}, \ion{Fe}{15}, and \ion{Fe}{16}. The blue circles correspond to pixels from Footpoint \#1 in Figure 1, while red crosses are from Footpoint \#2. The horizontal and vertical dashed lines denote the 3$\sigma$ level for each parameter. The $r$-values denote the correlation coefficient between the two parameters.}
\label{vel_nth_vel_fe14_15_16}
\end{center}
\end{figure*}

\begin{figure*}[]
\begin{center}
\includegraphics[height=16cm,angle=90]{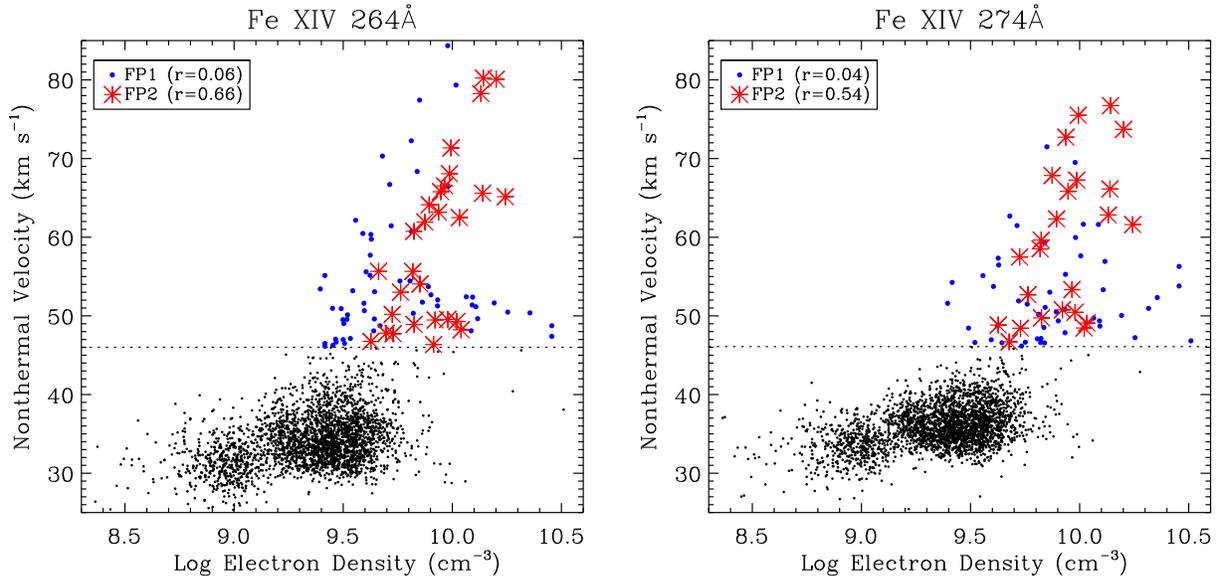}
\caption{Scatter plots of electron density against nonthermal velocity for the two \ion{Fe}{14} lines, 264\AA~and 274\AA. The horizontal and vertical dashed lines denote the 3$\sigma$ level for each parameter. The $r$-values denote the correlation coefficient between the two parameters.}
\label{den_nth_vel_fe14}
\end{center}
\end{figure*}

\begin{figure*}
\begin{center}
\includegraphics[height=16cm,angle=90]{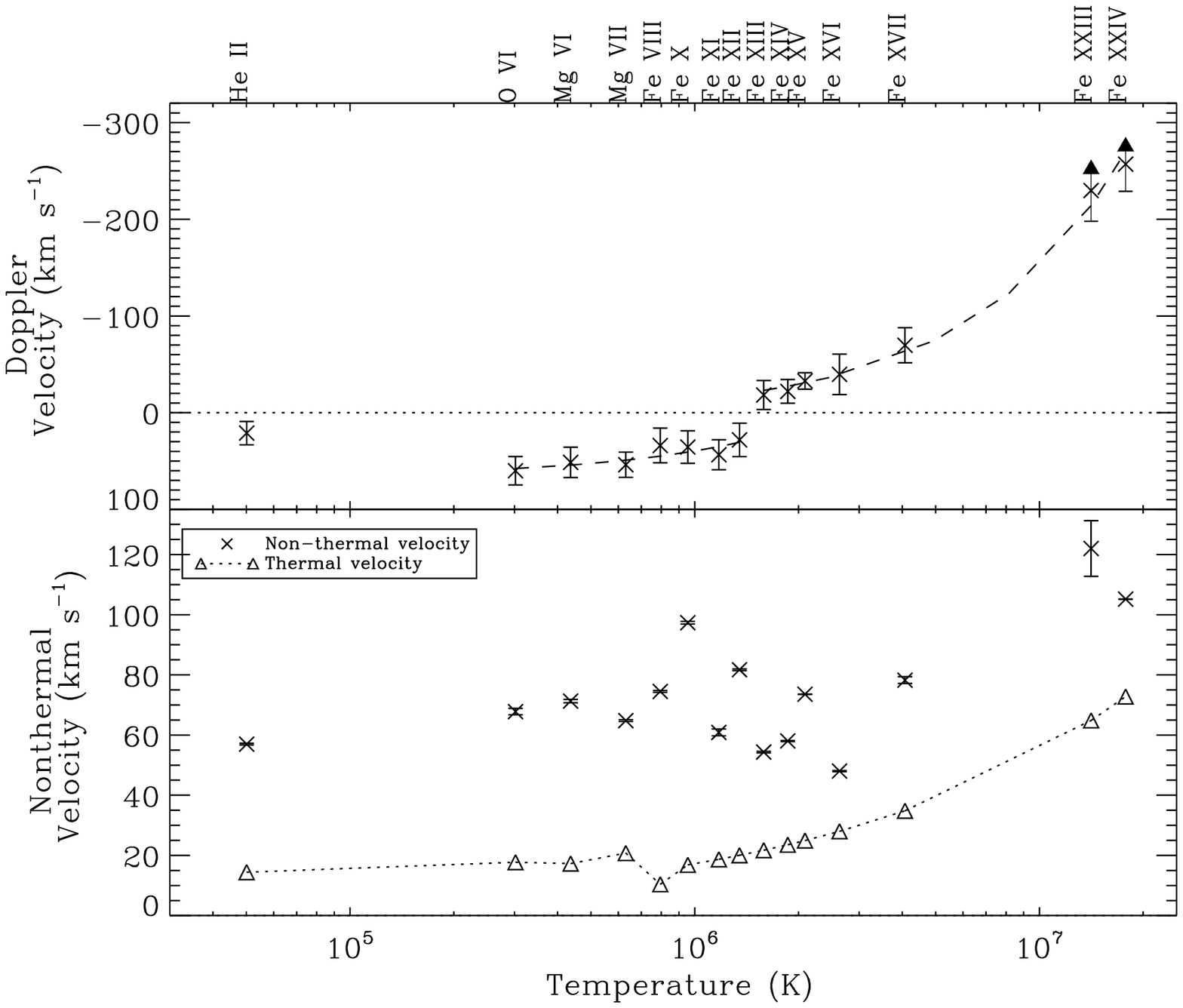}
\caption{Top panel: Line-of-sight Doppler velocity for the single brightest pixel in FP1 for each of the 15 emission lines studied. Positive velocities (redshifts) indicate downflows, while negative velocities (blueshifts) indicate upflows. Dashed lines represent least squares fits to the upward and downward moving plasma, excluding the \ion{He}{2} line (taken from \citealt{mill09}). Bottom panel: The corresponding nonthermal velocities for the same lines and at the same spatial location. The dotted line with triangles marks the thermal velocity calculated using Equation~\ref{eqn:therm_vel}. These values have already been removed from the nonthermal velocity calculations and this curve is merely for comparative purposes.}
\label{vel_nth_vel_temp_15}
\end{center}
\end{figure*}

\subsection{Nonthermal Velocity versus Doppler Velocity}
\label{vel_nth_vel}

Figure~\ref{vel_nth_vel_fe14_15_16} shows scatter plots of Doppler velocity against nonthermal velocity for the \ion{Fe}{14}, \ion{Fe}{15}, and \ion{Fe}{16} lines. The black data points centered around the 0~km~s$^{-1}$ level are from the quiescent active region and surrounding quiet Sun. The data points which are associated with the flaring emission from each footpoint are plotted as blue circles (FP1) and red crosses (FP2). It is shown that these values lie above the 3$\sigma$ level for each distribution as described at the beginning of Section~\ref{results}. While there appears to be a weak correlation between Doppler velocity and nonthermal velocity in each of these lines for FP1 ($\vert r \vert<0.39$, where $r$ is the Pearson correlation coefficient), the correlation between the two parameters for FP2 for the \ion{Fe}{15} and \ion{Fe}{16} lines is quite striking ($\vert r \vert>0.59$). There is a near-linear relationship between the two values indicating that, at least for this footpoint, that the broadening is a result of superposed Doppler flows which are due to heating by nonthermal electrons. From RHESSI observations it is known that nonthermal electrons have an energy distribution that closely resembles a power-law distribution. It is therefore reasonable to assume that this distribution of energies would translate to a broader range of velocities as it heats the lower layers of the atmosphere. This may result in the heated plasma becoming more turbulent, or in generating flows of evaporated material that are faster and slower than the bulk Doppler flow. The large degree of scatter for FP1 in each line could be due to the rastering nature of the observations: by the time the slit of the spectrometer had reached FP1 (rastering from right to left) the flare had become increasingly complex, with plasma flows sufficiently below the instrumental resolution.

\subsection{Nonthermal Velocity versus Electron Density}
\label{den_nth_vel}

The linear relationship between Doppler velocity and nonthermal velocity for FP2 derived in Section~\ref{vel_nth_vel} suggests that the excess broadening was due to unresolved plasma flows along the line of sight. To investigate whether the broadening could also be due to effects generated by the high densities obtained during the flare's impulsive phase, the nonthermal velocities for each of the two \ion{Fe}{14} lines (264\AA~and 274\AA) were plotted against the corresponding densities derived from the ratio of the two lines as described in Section~\ref{density}, and are shown in Figure~\ref{den_nth_vel_fe14}. These lines were the only lines available in the observing sequence that were both density sensitive and strong enough to derive reliable nonthermal velocities. 

Where Figure~\ref{vel_nth_vel_fe14_15_16} showed no discernible correlation between Doppler and nonthermal velocities for the \ion{Fe}{14} line, Figure~\ref{den_nth_vel_fe14} shows that there may be a stronger correlation between density and nonthermal velocity, at least for FP2 ($\vert r \vert >0.54$). FP1 on the other hand showed no distinguishable dependence between the two parameters ($\vert r \vert < 0.06$), with pixels which exhibited excessively high densities ($>$10$^{10}$~cm$^{-3}$) showing little or no sign of excess line broadening, and vice versa. This suggests that for FP2 at least (which was observed earlier in the flare than FP1) that the broadening of the \ion{Fe}{14} lines could have been due to pressure or opacity broadening because of the higher electron densities achieved during the initial heating phase. This conclusion is in contrast to that of \cite{dosc07} who found that regions of large line widths in active region studies did not correspond to regions of high density.

\subsection{Opacity Broadening or Pressure Broadening?}
\label{pressure_or_opacity}

To investigate whether either pressure or opacity effects might be the cause of the observed broadening in the \ion{Fe}{14} lines as deduced from Figure~\ref{den_nth_vel_fe14}, estimates can be made of how each of these effects contribute to the overall line profile. From \cite{bloo02} the opacity, $\tau_{0}$, can be estimated via:

\begin{equation}
\tau_{0} = 1.16 \times 10^{-14} \lambda f_{ij} \sqrt{\frac{M}{T}} \frac{n_{ion}}{n_{el}} \frac{n_{el}}{n_{H}} \frac{n_{H}}{N_{e}} N_{e}h
\label{opacity_eqn}
\end{equation}

\noindent
where $\lambda$ is the wavelength of the line, $f_{ij}$ is the oscillator strength (0.401 and 1.41 for the 264\AA~and 274\AA~lines, respectively; from \citealt{lian10}), $M$ is the mass of the ion (55.845 amu for Fe), $n_{Fe XIV}/n_{Fe} = 0.2$ (from \citealt{mazz98}), and $n_{Fe}/n_{H} = 10^{-4.49}$ (from \citealt{feld92}). Using these values, $\tau_{0}$ = 0.05 for the 264\AA~line and 0.2 for the 274\AA~line. Therefore both lines appear to be optically thin, which would suggest that opacity broadening was not significant.

So what about pressure broadening? For pressure broadening to be significant the collisional timescales have to be shorter than the timescale of the emitting photon, $t_{0}$, where $t_{0}$ is given by:

\begin{equation} 
\frac{1}{N_{e} \sigma \sqrt{2k_{B}T/M}}
\end{equation}

\noindent
where $N_{e}$ is the density and $\sigma$ is the collisional cross section of the ion. The expected amount of broadening is therefore:

\begin{equation}
\Delta \lambda = \frac{\lambda^{2}}{c} \frac{1}{\pi \Delta t_{0}} \approx \frac{\lambda^{2}}{c} \frac{N_{e} \sigma}{\pi} \sqrt{\frac{2k_{B}T}{M}}
\label{pressure_eqn}
\end{equation}

Taking $\sigma$ to be 5$\times$10$^{-19}$~cm$^{-2}$ (from \citealt{dere07}), $v_{th}$ = 58 km~s$^{-1}$ (from Table~\ref{line_data}), and a maximum density of 10$^{11}$~cm$^{-3}$, the effect of any pressure broadening equates to $\Delta \lambda$ $\approx$ 10$^{-15}$\AA, which is negligible in terms of nonthermal velocity. This therefore suggests than neither opacity nor pressure broadening alone can explain the density dependence on line widths as noted in Figure~\ref{den_nth_vel_fe14}.

\subsection{Doppler and Nonthermal Velocities as Functions of Temperature}
\label{vel_temp}

While it was not feasible to investigate the correlation between nonthermal velocity and electron density and velocity for other lines due to poor signal-to-noise ratios, as seen in the bottom row of Figure~\ref{eis_int_vel_wid_maps}, and the lack of appropriate density sensitive line ratios, the nonthermal velocity at the brightest footpoint pixel in the raster (in FP1) was measurable for lines formed over the broad range of temperatures. It was from this pixel that \cite{mill09} determined the linear relationship between Doppler velocity and temperature. Figure~\ref{vel_nth_vel_temp_15} shows these results in addition to the corresponding nonthermal velocities for the same lines plotted against the formation temperature of the line. Also plotted are the values of the thermal velocities for each line (dashed line with triangles) calculated from Equation~\ref{eqn:therm_vel} using the formation temperatures listed in Table~\ref{line_data}. (Note that the thermal width has already been removed from the total line width before calculating the nonthermal velocity; this curve merely acts as a comparative guide for the values of the thermal velocities for each line.) The coolest line in the observing sequence, \ion{He}{2}, displayed a nonthermal velocity of $\sim$55~km~s$^{-1}$ while the hottest lines (\ion{Fe}{23} and \ion{Fe}{24}) showed values greater than 100~km~s$^{-1}$. However, care must be taken when evaluating the magnitude of the widths for these lines as the \ion{He}{0} line is known to be blended with \ion{Si}{10}, \ion{Fe}{12}, and \ion{Fe}{13} \citep{youn07}, and both the blueshifted components of the \ion{Fe}{23} and \ion{Fe}{24} lines were measured near the edges of their respective spectral windows (see Figure 4 in \citealt{mill09}), so the resulting Gaussian fits may not be wholly accurate. The lack of a systematic correlation between nonthermal velocity and temperature, as found with Doppler velocities, suggests that the line broadening may not be solely due to a superposition of plasma flows below the instrumental resolution.

\section{Conclusions}
\label{conc}

This paper presents a detailed investigation into the nature of spatially-resolved line broadening of EUV emission lines during the impulsive phase of a C-class solar flare. Line profiles, co-spatial with the HXR emission observed by RHESSI, were found to be broadened beyond their thermal widths. Using techniques similar to that used to establish the cause of line broadening in quiescent active region spectra \citep{hara08,dosc08,brya10,pete10}, it was found that a strong correlation existed between Doppler velocity and nonthermal velocity for the \ion{Fe}{15} and \ion{Fe}{16} lines at one of the footpoints. This suggests that the line broadening at these temperatures was a signature of unresolved plasma flows along the line of sight during the process of chromospheric evaporation by nonthermal electrons.

The analysis of the \ion{Fe}{14} line on the other hand, which showed no conclusive correlation between Doppler and nonthermal velocities,
showed a stronger correlation between electron density and nonthermal velocity which suggested that the excess line broadening at these temperatures cold have been due to either opacity or pressure broadening. However, estimates of the magnitude of each of these effects appeared to suggest that the amount of excess broadening was negligible in each case. Perhaps the assumptions made in solving Equations~\ref{opacity_eqn} and \ref{pressure_eqn} were incorrect (e.g. ionization equilibrium; see below), or the broadening was due to a culmination of different effects, or perhaps it was due to a different mechanism altogether not considered here (e.g. Stark broadening). While the findings presented here suggest tentative evidence for line broadening due to enhanced electron densities during a C-class flare, perhaps larger, more energetic events, or density diagnostics of higher temperature plasmas, will show these effects to be even more substantial. Line broadening can not only reveal important information with regard to the heating processes during flares but can also be a crucial diagnostic of the fundamental atomic physics and must be a component of future flare modelling. 

The underlying assumption of this analysis was that the lines investigated were formed in ionization equilibrium. While this assumption is usually valid for high-density plasmas \citep{brad10}, departures from equilibrium can affect the assumed formation temperature of a line. If a line was formed at a higher temperature than that quoted in Table~\ref{line_data}, then the resulting nonthermal velocity could be much less than measured here, perhaps even negligible. For example, the nonthermal velocity calculated for the \ion{Fe}{15} line was 73~km~s~$^{-1}$. At the assumed formation temperature of 2~MK this yields a thermal velocity of 25~km~s$^{-1}$. If the formation temperature was increased to $\sim$8~MK then the nonthermal width would essentially tend to zero. However, this would also result in a decrease in the line intensity by three orders of magnitude as determined by the corresponding contribution function.

While previous studies of emission line widths during solar flares have often focused on line profiles integrated over the entire solar disk, EIS now offers the capability of determining the location and magnitude of the broadening thanks to its superior spectral resolution. This, coupled with its remarkable Doppler resolution, density diagnostic capability, and broad temperature coverage allow a truly detailed study of the composition and dynamic behavior of the flaring solar atmosphere.

\acknowledgments
The author would like to thank Peter Young for his assistance with the density diagnostics and for feedback on the manuscript, Brian Dennis and Gordon Holman for their insightful and stimulating discussions, Mihalis Mathioudakis and Francis Keenan for discussions on opacity, the anonymous referee for their constructive comments, the International Space Science Institute (ISSI, Bern) for the opportunity to discuss these results at the international team meeting on chromospheric flares, and Queen's University Belfast for the award of a Leverhulme Trust Research Fellowship. Hinode is a Japanese mission developed and launched by ISAS/JAXA, collaborating with NAOJ as domestic partner, and NASA (USA) and STFC (UK) as international partners. Scientific operation of the Hinode mission is conducted by the Hinode science team organized at ISAS/JAXA. This team mainly consists of scientists from institutes in the partner countries. Support for the post-launch operation is provided by JAXA and NAOJ, STFC, NASA, ESA (European Space Agency), and NSC (Norway).


\begin{thebibliography}{58}
\expandafter\ifx\csname natexlab\endcsname\relax\def\natexlab#1{#1}\fi

\bibitem[{REV(????)}]{REVTEX41Control}
 ????

\bibitem[{08(1)}]{apsrev41Control}
08. 1

\bibitem[{{Acton} {et~al.}(1982){Acton}, {Leibacher}, {Canfield}, {Gunkler},
  {Hudson}, \& {Kiplinger}}]{acto82}
{Acton}, L.~W., {Leibacher}, J.~W., {Canfield}, R.~C., {Gunkler}, T.~A.,
  {Hudson}, H.~S., \& {Kiplinger}, A.~L. 1982, \apj, 263, 409

\bibitem[{{Acton} {et~al.}(1980){Acton}, {Finch}, {Gilbreth}, {Culhane},
  {Bentley}, {Bowles}, {Guttridge}, {Gabriel}, {Firth}, \& {Hayes}}]{acto81}
{Acton}, L.~W., {et~al.} 1980, \solphys, 65, 53

\bibitem[{{Antiochos} \& {Sturrock}(1982)}]{anto82}
{Antiochos}, S.~K., \& {Sturrock}, P.~A. 1982, \apj, 254, 343

\bibitem[{{Antonucci} \& {Dennis}(1983)}]{anto83}
{Antonucci}, E., \& {Dennis}, B.~R. 1983, \solphys, 86, 67

\bibitem[{{Bloomfield} {et~al.}(2002){Bloomfield}, {Mathioudakis}, {Christian},
  {Keenan}, \& {Linsky}}]{bloo02}
{Bloomfield}, D.~S., {Mathioudakis}, M., {Christian}, D.~J., {Keenan}, F.~P.,
  \& {Linsky}, J.~L. 2002, \aap, 390, 219

\bibitem[{{Bradshaw} \& {Cargill}(2010)}]{brad10}
{Bradshaw}, S.~J., \& {Cargill}, P.~J. 2010, \apj, 717, 163

\bibitem[{{Brosius}(2009)}]{bros09b}
{Brosius}, J.~W. 2009, \apj, 701, 1209

\bibitem[{{Brosius} \& {Holman}(2007)}]{bros07}
{Brosius}, J.~W., \& {Holman}, G.~D. 2007, \apjl, 659, L73

\bibitem[{{Brosius} \& {Holman}(2009)}]{bros09a}
---. 2009, \apj, 692, 492

\bibitem[{{Brosius} \& {Phillips}(2004)}]{bros04}
{Brosius}, J.~W., \& {Phillips}, K.~J.~H. 2004, \apj, 613, 580

\bibitem[{{Bryans} {et~al.}(2010){Bryans}, {Young}, \& {Doschek}}]{brya10}
{Bryans}, P., {Young}, P.~R., \& {Doschek}, G.~A. 2010, \apj, 715, 1012

\bibitem[{{Canfield} {et~al.}(1984){Canfield}, {Gunkler}, \&
  {Ricchiazzi}}]{canf84}
{Canfield}, R.~C., {Gunkler}, T.~A., \& {Ricchiazzi}, P.~J. 1984, \apj, 282,
  296

\bibitem[{{Canfield} {et~al.}(1987){Canfield}, {Metcalf}, {Strong}, \&
  {Zarro}}]{canf87}
{Canfield}, R.~C., {Metcalf}, T.~R., {Strong}, K.~T., \& {Zarro}, D.~M. 1987,
  \nat, 326, 165

\bibitem[{{Chifor} {et~al.}(2008){Chifor}, {Young}, {Isobe}, {Mason},
  {Tripathi}, {Hara}, \& {Yokoyama}}]{chif08}
{Chifor}, C., {Young}, P.~R., {Isobe}, H., {Mason}, H.~E., {Tripathi}, D.,
  {Hara}, H., \& {Yokoyama}, T. 2008, \aap, 481, L57

\bibitem[{{Christian} {et~al.}(2004){Christian}, {Mathioudakis}, {Bloomfield},
  {Dupuis}, \& {Keenan}}]{chri04}
{Christian}, D.~J., {Mathioudakis}, M., {Bloomfield}, D.~S., {Dupuis}, J., \&
  {Keenan}, F.~P. 2004, \apj, 612, 1140

\bibitem[{{Christian} {et~al.}(2006){Christian}, {Mathioudakis}, {Bloomfield},
  {Dupuis}, {Keenan}, {Pollacco}, \& {Malina}}]{chri06}
{Christian}, D.~J., {Mathioudakis}, M., {Bloomfield}, D.~S., {Dupuis}, J.,
  {Keenan}, F.~P., {Pollacco}, D.~L., \& {Malina}, R.~F. 2006, \aap, 454, 889

\bibitem[{{Culhane} {et~al.}(1991){Culhane}, {Bentley}, {Hiei}, {Watanabe},
  {Doschek}, {Brown}, {Cruise}, {Lang}, {Ogawara}, \& {Uchida}}]{culh91}
{Culhane}, J.~L., {et~al.} 1991, \solphys, 136, 89

\bibitem[{{Czaykowska} {et~al.}(2001){Czaykowska}, {Alexander}, \& {De
  Pontieu}}]{czay01}
{Czaykowska}, A., {Alexander}, D., \& {De Pontieu}, B. 2001, \apj, 552, 849

\bibitem[{{Czaykowska} {et~al.}(1999){Czaykowska}, {de Pontieu}, {Alexander},
  \& {Rank}}]{czay99}
{Czaykowska}, A., {de Pontieu}, B., {Alexander}, D., \& {Rank}, G. 1999, \apjl,
  521, L75

\bibitem[{{Del Zanna} {et~al.}(2011){Del Zanna}, {Mitra-Kraev}, {Bradshaw},
  {Mason}, \& {Asai}}]{delz11}
{Del Zanna}, G., {Mitra-Kraev}, U., {Bradshaw}, S.~J., {Mason}, H.~E., \&
  {Asai}, A. 2011, \aap, 526, A1+

\bibitem[{{Dere}(2007)}]{dere07}
{Dere}, K.~P. 2007, \aap, 466, 771

\bibitem[{{Dere} {et~al.}(2009){Dere}, {Landi}, {Young}, {Del Zanna},
  {Landini}, \& {Mason}}]{dere09}
{Dere}, K.~P., {Landi}, E., {Young}, P.~R., {Del Zanna}, G., {Landini}, M., \&
  {Mason}, H.~E. 2009, \aap, 498, 915

\bibitem[{{Doschek} {et~al.}(1980){Doschek}, {Feldman}, {Kreplin}, \&
  {Cohen}}]{dosc80}
{Doschek}, G.~A., {Feldman}, U., {Kreplin}, R.~W., \& {Cohen}, L. 1980, \apj,
  239, 725

\bibitem[{{Doschek} {et~al.}(2007){Doschek}, {Mariska}, {Warren}, {Culhane},
  {Watanabe}, {Young}, {Mason}, \& {Dere}}]{dosc07}
{Doschek}, G.~A., {Mariska}, J.~T., {Warren}, H.~P., {Culhane}, L., {Watanabe},
  T., {Young}, P.~R., {Mason}, H.~E., \& {Dere}, K.~P. 2007, \pasj, 59, 707

\bibitem[{{Doschek} \& {Warren}(2005)}]{dosc05}
{Doschek}, G.~A., \& {Warren}, H.~P. 2005, \apj, 629, 1150

\bibitem[{{Doschek} {et~al.}(2008){Doschek}, {Warren}, {Mariska}, {Muglach},
  {Culhane}, {Hara}, \& {Watanabe}}]{dosc08}
{Doschek}, G.~A., {Warren}, H.~P., {Mariska}, J.~T., {Muglach}, K., {Culhane},
  J.~L., {Hara}, H., \& {Watanabe}, T. 2008, \apj, 686, 1362

\bibitem[{{Feldman}(1992)}]{feld92}
{Feldman}, U. 1992, \physscr, 46, 202

\bibitem[{{Feldman} {et~al.}(1980){Feldman}, {Doschek}, {Kreplin}, \&
  {Mariska}}]{feld80}
{Feldman}, U., {Doschek}, G.~A., {Kreplin}, R.~W., \& {Mariska}, J.~T. 1980,
  \apj, 241, 1175

\bibitem[{{Gabriel} {et~al.}(1981){Gabriel}, {Phillips}, {Acton}, {Wolfson},
  {Culhane}, {Rapley}, {Bentley}, {Kayat}, {Jordan}, \& {Antonucci}}]{gabr81}
{Gabriel}, A.~H., {et~al.} 1981, \apjl, 244, L147

\bibitem[{{Gallagher} {et~al.}(2001){Gallagher}, {Phillips}, {Lee}, {Keenan},
  \& {Pinfield}}]{gall01}
{Gallagher}, P.~T., {Phillips}, K.~J.~H., {Lee}, J., {Keenan}, F.~P., \&
  {Pinfield}, D.~J. 2001, \apj, 558, 411

\bibitem[{{Graham} {et~al.}(2011){Graham}, {Fletcher}, \& {Hannah}}]{grah11}
{Graham}, D.~R., {Fletcher}, L., \& {Hannah}, I.~G. 2011, \aap, Submitted

\bibitem[{{Grineva} {et~al.}(1973){Grineva}, {Karev}, {Korneev}, {Krutov},
  {Mandelstam}, {Vainstein}, {Vasilyev}, \& {Zhitnik}}]{grin73}
{Grineva}, Y.~I., {Karev}, V.~I., {Korneev}, V.~V., {Krutov}, V.~V.,
  {Mandelstam}, S.~L., {Vainstein}, L.~A., {Vasilyev}, B.~N., \& {Zhitnik},
  I.~A. 1973, \solphys, 29, 441

\bibitem[{{Handy} {et~al.}(1999){Handy}, {Acton}, {Kankelborg}, {Wolfson},
  {Akin}, {Bruner}, {Caravalho}, {Catura}, {Chevalier}, {Duncan}, {Edwards},
  {Feinstein}, {Freeland}, {Friedlaender}, {Hoffmann}, {Hurlburt}, {Jurcevich},
  {Katz}, {Kelly}, {Lemen}, {Levay}, {Lindgren}, {Mathur}, {Meyer}, {Morrison},
  {Morrison}, {Nightingale}, {Pope}, {Rehse}, {Schrijver}, {Shine}, {Shing},
  {Strong}, {Tarbell}, {Title}, {Torgerson}, {Golub}, {Bookbinder}, {Caldwell},
  {Cheimets}, {Davis}, {Deluca}, {McMullen}, {Warren}, {Amato}, {Fisher},
  {Maldonado}, \& {Parkinson}}]{hand99}
{Handy}, B.~N., {et~al.} 1999, \solphys, 187, 229

\bibitem[{{Hara} {et~al.}(2009){Hara}, {Watanabe}, {Bone}, {Culhane}, {van
  Driel-Gesztelyi}, \& {Young}}]{hara09}
{Hara}, H., {Watanabe}, T., {Bone}, L.~A., {Culhane}, J.~L., {van
  Driel-Gesztelyi}, L., \& {Young}, P.~R. 2009, in Astronomical Society of the
  Pacific Conference Series, Vol. 415, Astronomical Society of the Pacific
  Conference Series, ed. {B.~Lites, M.~Cheung, T.~Magara, J.~Mariska, \&
  K.~Reeves}, 459--+

\bibitem[{{Hara} {et~al.}(2008){Hara}, {Watanabe}, {Harra}, {Culhane}, {Young},
  {Mariska}, \& {Doschek}}]{hara08}
{Hara}, H., {Watanabe}, T., {Harra}, L.~K., {Culhane}, J.~L., {Young}, P.~R.,
  {Mariska}, J.~T., \& {Doschek}, G.~A. 2008, \apjl, 678, L67

\bibitem[{{Harra} {et~al.}(2009){Harra}, {Williams}, {Wallace}, {Magara},
  {Hara}, {Tsuneta}, {Sterling}, \& {Doschek}}]{harr09}
{Harra}, L.~K., {Williams}, D.~R., {Wallace}, A.~J., {Magara}, T., {Hara}, H.,
  {Tsuneta}, S., {Sterling}, A.~C., \& {Doschek}, G.~A. 2009, \apjl, 691, L99

\bibitem[{{Harrison} {et~al.}(1995){Harrison}, {Sawyer}, {Carter}, {Cruise},
  {Cutler}, {Fludra}, {Hayes}, {Kent}, {Lang}, {Parker}, {Payne}, {Pike},
  {Peskett}, {Richards}, {Gulhane}, {Norman}, {Breeveld}, {Breeveld}, {Al
  Janabi}, {McCalden}, {Parkinson}, {Self}, {Thomas}, {Poland}, {Thomas},
  {Thompson}, {Kjeldseth-Moe}, {Brekke}, {Karud}, {Maltby}, {Aschenbach},
  {Br{\"a}uninger}, {K{\"u}hne}, {Hollandt}, {Siegmund}, {Huber}, {Gabriel},
  {Mason}, \& {Bromage}}]{harr95}
{Harrison}, R.~A., {et~al.} 1995, \solphys, 162, 233

\bibitem[{{Imada} {et~al.}(2008){Imada}, {Hara}, {Watanabe}, {Asai},
  {Minoshima}, {Harra}, \& {Mariska}}]{imad08}
{Imada}, S., {Hara}, H., {Watanabe}, T., {Asai}, A., {Minoshima}, T., {Harra},
  L.~K., \& {Mariska}, J.~T. 2008, \apjl, 679, L155

\bibitem[{{Johns-Krull} {et~al.}(1997){Johns-Krull}, {Hawley}, {Basri}, \&
  {Valenti}}]{john97}
{Johns-Krull}, C.~M., {Hawley}, S.~L., {Basri}, G., \& {Valenti}, J.~A. 1997,
  \apjs, 112, 221

\bibitem[{{Lee} {et~al.}(1996){Lee}, {Lee}, {Yun}, {Fang}, \& {Hu}}]{lee96}
{Lee}, S., {Lee}, J., {Yun}, H.~S., {Fang}, C., \& {Hu}, J. 1996, \apjl, 470,
  L65+

\bibitem[{{Liang} {et~al.}(2010){Liang}, {Badnell}, {Crespo L{\'o}pez-Urrutia},
  {Baumann}, {Del Zanna}, {Storey}, {Tawara}, \& {Ullrich}}]{lian10}
{Liang}, G.~Y., {Badnell}, N.~R., {Crespo L{\'o}pez-Urrutia}, J.~R., {Baumann},
  T.~M., {Del Zanna}, G., {Storey}, P.~J., {Tawara}, H., \& {Ullrich}, J. 2010,
  \apjs, 190, 322

\bibitem[{{Lin} {et~al.}(2002){Lin}, {Dennis}, {Hurford}, {Smith}, {Zehnder},
  {Harvey}, {Curtis}, {Pankow}, {Turin}, {Bester}, {Csillaghy}, {Lewis},
  {Madden}, {van Beek}, {Appleby}, {Raudorf}, {McTiernan}, {Ramaty}, {Schmahl},
  {Schwartz}, {Krucker}, {Abiad}, {Quinn}, {Berg}, {Hashii}, {Sterling},
  {Jackson}, {Pratt}, {Campbell}, {Malone}, {Landis}, {Barrington-Leigh},
  {Slassi-Sennou}, {Cork}, {Clark}, {Amato}, {Orwig}, {Boyle}, {Banks},
  {Shirey}, {Tolbert}, {Zarro}, {Snow}, {Thomsen}, {Henneck}, {McHedlishvili},
  {Ming}, {Fivian}, {Jordan}, {Wanner}, {Crubb}, {Preble}, {Matranga}, {Benz},
  {Hudson}, {Canfield}, {Holman}, {Crannell}, {Kosugi}, {Emslie}, {Vilmer},
  {Brown}, {Johns-Krull}, {Aschwanden}, {Metcalf}, \& {Conway}}]{lin02}
{Lin}, R.~P., {et~al.} 2002, \solphys, 210, 3

\bibitem[{{Mazzotta} {et~al.}(1998){Mazzotta}, {Mazzitelli}, {Colafrancesco},
  \& {Vittorio}}]{mazz98}
{Mazzotta}, P., {Mazzitelli}, G., {Colafrancesco}, S., \& {Vittorio}, N. 1998,
  \aaps, 133, 403

\bibitem[{{Milligan}(2008)}]{mill08}
{Milligan}, R.~O. 2008, \apjl, 680, L157

\bibitem[{{Milligan} \& {Dennis}(2009)}]{mill09}
{Milligan}, R.~O., \& {Dennis}, B.~R. 2009, \apj, 699, 968

\bibitem[{{Milligan} {et~al.}(2006{\natexlab{a}}){Milligan}, {Gallagher},
  {Mathioudakis}, {Bloomfield}, {Keenan}, \& {Schwartz}}]{mill06a}
{Milligan}, R.~O., {Gallagher}, P.~T., {Mathioudakis}, M., {Bloomfield}, D.~S.,
  {Keenan}, F.~P., \& {Schwartz}, R.~A. 2006{\natexlab{a}}, \apjl, 638, L117

\bibitem[{{Milligan} {et~al.}(2006{\natexlab{b}}){Milligan}, {Gallagher},
  {Mathioudakis}, \& {Keenan}}]{mill06b}
{Milligan}, R.~O., {Gallagher}, P.~T., {Mathioudakis}, M., \& {Keenan}, F.~P.
  2006{\natexlab{b}}, \apjl, 642, L169

\bibitem[{{Milligan} {et~al.}(2005){Milligan}, {Gallagher}, {Mathioudakis},
  {Keenan}, \& {Bloomfield}}]{mill05}
{Milligan}, R.~O., {Gallagher}, P.~T., {Mathioudakis}, M., {Keenan}, F.~P., \&
  {Bloomfield}, D.~S. 2005, \mnras, 363, 259

\bibitem[{{Peter}(2010)}]{pete10}
{Peter}, H. 2010, \aap, 521, A51+

\bibitem[{{Saint-Hilaire} {et~al.}(2010){Saint-Hilaire}, {Krucker}, \&
  {Lin}}]{sain10}
{Saint-Hilaire}, P., {Krucker}, S., \& {Lin}, R.~P. 2010, \apj, 721, 1933

\bibitem[{{Teriaca} {et~al.}(2003){Teriaca}, {Falchi}, {Cauzzi}, {Falciani},
  {Smaldone}, \& {Andretta}}]{teri03}
{Teriaca}, L., {Falchi}, A., {Cauzzi}, G., {Falciani}, R., {Smaldone}, L.~A.,
  \& {Andretta}, V. 2003, \apj, 588, 596

\bibitem[{{Warren} \& {Winebarger}(2003)}]{warr03}
{Warren}, H.~P., \& {Winebarger}, A.~R. 2003, \apjl, 596, L113

\bibitem[{{Young}(2011)}]{youn11}
{Young}, P.~R. 2011, EIS Software Note 15

\bibitem[{{Young} {et~al.}(2009){Young}, {Watanabe}, {Hara}, \&
  {Mariska}}]{youn09}
{Young}, P.~R., {Watanabe}, T., {Hara}, H., \& {Mariska}, J.~T. 2009, \aap,
  495, 587

\bibitem[{{Young} {et~al.}(2007){Young}, {Del Zanna}, {Mason}, {Dere}, {Landi},
  {Landini}, {Doschek}, {Brown}, {Culhane}, {Harra}, {Watanabe}, \&
  {Hara}}]{youn07}
{Young}, P.~R., {et~al.} 2007, \pasj, 59, 857

\bibitem[{{Zarro} \& {Lemen}(1988)}]{zarr88}
{Zarro}, D.~M., \& {Lemen}, J.~R. 1988, \apj, 329, 456

\end{thebibliography}
\end{document}